\documentclass[aps,preprint,floatfix,superscriptaddress]{revtex4-2}
\usepackage{amsmath,amssymb,amsfonts}
\usepackage{graphicx}
\usepackage[svgnames]{xcolor}
\usepackage[colorlinks=True,linkcolor=DarkRed,citecolor=ForestGreen,urlcolor=MediumBlue,pdfstartview=FitH,bookmarks=False,pdfpagemode=UseNone]{hyperref}
\usepackage{newtxtext,newtxmath}
\usepackage{microtype}
\usepackage{bm}\let\vec\bm

\def\kTF{k_{\mathrm{s}}}
\def\eF{\varepsilon_{\mathrm{F}}}
\def\aB{a_0}
\def\kF{k_{\mathrm{F}}}
\def\kB{k_{\mathrm{B}}}
\def\Tc{T_{\mathrm{c}}}
\def\ne{n_{\mathrm{e}}}
\def\me{m_{\mathrm{e}}}
\def\mp{m_{\mathrm{p}}}
\def\wpn{\omega_{\mathrm{s}}}
\def\Epn{\Omega_{\mathrm{s}}}
\def\cpn{c_{\mathrm{s}}}

\begin{document}

\title{Drosophila of phonon-mediated superconductivity: Full Eliashberg theory of the jellium model}

\author{Christophe Berthod}
\affiliation{Department of Quantum Matter Physics, University of Geneva, 24 quai Ernest-Ansermet, 1211 Geneva, Switzerland}

\author{Louk Rademaker}
\affiliation{Department of Quantum Matter Physics, University of Geneva, 24 quai Ernest-Ansermet, 1211 Geneva, Switzerland}
\affiliation{Lorentz Institute for Theoretical Physics, Leiden University, PO Box 9506, 2300 Leiden, The Netherlands}

\author{Dirk van der Marel}
\affiliation{Department of Quantum Matter Physics, University of Geneva, 24 quai Ernest-Ansermet, 1211 Geneva, Switzerland}

\date{\today}

\begin{abstract}

We present a full numerical solution of the Migdal--Eliashberg equations for the jellium model of phonon-mediated superconductivity. We find very low critical temperatures below 1~K, in contrast to earlier claims that the jellium model for hydrogen solids might reach room-temperature superconductivity. Our results suggest that full momentum and frequency dependence of the gap function and normal self-energy should be taken into account for accurate $\Tc$ estimates.

\end{abstract}

\maketitle

\section{Introduction}

The sixty-four-thousand-dollar question of high-$\Tc$ superconductivity is why the transition temperature is so high---and how to get it higher. An answer to this question should point to a specific \emph{mechanism}, and over the years a true cornucopia of mechanisms have been theoretically developed.

The main challenge of any mechanism is that it requires attraction between electrons, even though electrons obviously repel each other through the Coulomb interaction. One can overcome this overall repulsion with a sign change of the order parameter. In $d$-wave superconductors, this sign change is clearly visible in momentum space. The sign change in phonon-mediated superconductivity is more subtle, however. After all, even the electron-phonon coupling itself is derived from the Coulomb interaction! To overcome the strong repulsion, the order parameter changes sign as a function of \emph{frequency}. This is the famous retardation effect, driven by the fact that the interactions mediated by phonons are slower than the instantaneous Coulomb repulsion. An interesting perspective on this was provided by Chester \cite{chester1956difference}, who demonstrated on the basis of the virial theorem that the total Coulomb interaction of all electrons and nuclei together should be smaller in the superconducting state than in the normal state.

The possibility of superconductivity is therefore completely determined by the full momentum and frequency dependence of the screened Coulomb interaction,
\begin{equation}
	V_{\mathrm{eff}}(\vec{q}, \omega) = \frac{e^2}{\epsilon_0 q^2 \epsilon(\vec{q}, \omega)},
	\label{Eq:EffectiveInteraction}
\end{equation}
where $\epsilon(\vec{q}, \omega)$ is the dielectric function of the system of interest and $q=|\vec{q}|$. The question of the mechanism then reverts to the question of establishing the correct form of the dielectric function.

Among the long list of mechanisms, one stands out due to its elegance and historical significance: the jellium model \cite{degennes1966superconductivity}. The concept is simple: rather than boring us with microscopic details, the jellium model assumes the electron gas lives inside a homogeneous charge-compensating medium. The background medium has sound waves (phonons) corresponding to density fluctuations, whose dispersion is defined through
\begin{equation}
	\omega_{\mathrm{ph}}^2(\vec{q}) = \frac{\wpn^2q^2}{\kTF^2 + q^2}
\end{equation}
where $\kTF$ is the inverse Thomas--Fermi screening length given by
\begin{equation}
	\kTF^2 = \frac{e^2 \me}{\pi^{4/3} \epsilon_0 \hbar^2}\, (3{\ne})^{1/3}
\end{equation}
with $\me$ the electron mass and $\ne$ the electron density. The phonon dispersion is linear for long wavelengths with velocity $\wpn/\kTF$. At large momenta, the dispersion tends to the plasma frequency $\wpn$ of the background medium defined through
\begin{equation}
	\wpn^2 = \frac{{\ne} Z_{\mathrm{I}} e^2}{\epsilon_0 M},
	\label{Eq:PlasmaFrequency}
\end{equation}
where $M$ is the mass and $Z_{\mathrm{I}}$ the charge of the ions making up the jellium. The resulting jellium dielectric function is given by two terms \cite{bardeen1955electron,morel1962calculation}: the static Thomas--Fermi screened Coulomb plus the phonon contribution,
\begin{equation}
	\frac{1}{\epsilon(\vec{q},\omega)} = \frac{q^2}{\kTF^2 + q^2}
	\left[ 1 + \frac{\omega_{\mathrm{ph}}^2(\vec{q})}{\omega^2 - \omega_{\mathrm{ph}}^2(\vec{q})} \right].
\end{equation}
In the present paper we will \emph{not} make this separation. Instead we will treat the Coulomb repulsion and the phonon-mediated interaction on the same footing. This simplifies the dielectric function to
\begin{equation}
	\epsilon(\vec{q},\omega)=1+\frac{\kTF^2}{q^2}-\left(\frac{\wpn}{\omega}\right)^2,
	\label{Eq:JelliumDielectric}
\end{equation}
which defines, together with Eq.~\eqref{Eq:EffectiveInteraction}, the electron-electron interaction of the jellium model. Despite its simplicity, the jellium model contains all the ingredients necessary for superconductivity. Most importantly, the static repulsion is overcome for small frequencies: in the range $\omega < \omega_{\mathrm{ph}}(\vec{q})$, mediation by the phonon turns the overall interaction attractive. Thanks to this retardation, the jellium model is the drosophila of phonon-mediated superconductivity.

The plasma frequency of Eq.~\eqref{Eq:PlasmaFrequency} sets an upper bound on the energy of the phonon; and plays the role of the Debye frequency in regular BCS theory \cite{bardeen1957theory}. Naively, the highest $\Tc$ can be obtained by maximizing this frequency, which would require taking the smallest possible mass for the ions. That is the origin of the quest for hydrogen-based superconductivity, with $M = \mp$, where $\mp$ is the proton mass. Indeed, plugging the hydrogen mass into BCS-like expressions gives room-temperature superconductivity. The question is whether this result survives the full effects of retardation and repulsion.

Recently, two of us followed in the footsteps of Ginzburg and Kirzhnits \cite{ginzburg1968superconductivity, kirzhnits1969superconductivity} and calculated the momentum dependence of the gap function and $\Tc$ using the ``on-energy-shell'' implementation of the jellium model, obtaining a non-trivial superconducting solution with $\Tc$ strongly depending on density with an upper bound of only 30~K \cite{marel2024superconductivity}. This approach, which amounts to substituting the frequency dependence of the dielectric function with a momentum exchange, $\hbar\omega =\varepsilon_{\vec{k}+\vec{q}}-\varepsilon_{\vec{k}}$, has the drawback that it doesn't do justice to the aspect of retardation. The solution of the jellium model taking into account retardation effects \emph{and} the full momentum dependence of the interaction \cite{degennes1966superconductivity} has, until recently, not been reported in the literature. An important improvement in this respect was presented in a recent preprint \cite{sadovskii2026} using the Kirzhnits--Maksimov--Khomskii (KMK) \cite{kirzhnits1973the} formalism, resulting in a much lower upper bound of 0.05~K. This formalism takes into account both retardation and momentum dependence of the interaction but, in contrast to the Migdal--Eliashberg approach \cite{eliashberg1960interactions, parks1969superconductivity, marsiglio2020eliashberg}, it is limited to weak coupling. As the previous estimates of the superconducting $\Tc$ fail to take explicitly into account retardation through the Migdal--Eliashberg approach \cite{eliashberg1960interactions, parks1969superconductivity, marsiglio2020eliashberg}, their reliability is questionable. To address this question, in this work we solve the Migdal--Eliashberg equation in its full momentum- and frequency-dependent form for the jellium model. We find that the hydrogen-based jellium model does not lead to superconducting temperatures in excess of $\Tc \approx 1$~K. So, while the model is simple, the jellium fruit fly seems not to be a good direction to search for high-$\Tc$ superconductivity.

\section{The Migdal--Eliashberg equations}

In BCS theory, the superconducting phase is characterized by a nonzero \emph{static} Cooper pair expectation value $\langle c^\dagger_{\vec{k} \sigma} c^\dagger_{-\vec{k} \sigma'}\rangle$, determined through a mean-field equation. In Migdal--Eliashberg (ME) theory \cite{eliashberg1960interactions, parks1969superconductivity, marsiglio2020eliashberg}, this is elevated to a \emph{dynamical} quantity called the anomalous self-energy. The ME equations are most conveniently expressed in the Nambu spinor basis,
\begin{equation}
	\Psi_{\vec{k}} = \begin{pmatrix}
		c_{\vec{k} \uparrow} \\ c^\dagger_{-\vec{k} \downarrow}
	\end{pmatrix},
\end{equation}
combining the electron with spin up at wavevector $\vec{k}$ with the hole with spin down at wavevector $-\vec{k}$. The imaginary-time Nambu--Gorkov Green's function is now a matrix (indicated by a hat) in spinor space,
\begin{equation}
	\hat{G}(\vec{k},\tau) = -\langle T_\tau \Psi^{\phantom\dagger}_{\vec{k}} (\tau) \Psi^\dagger_{\vec{k}} \rangle
	= \begin{pmatrix}
		G(\vec{k},\tau) & F(\vec{k},\tau) \\
		F(\vec{k},\tau) & -G(\vec{k},-\tau)
	\end{pmatrix},
\end{equation}
where $G(\vec{k},\tau) = - \langle T_\tau c^{\phantom\dagger}_{\vec{k} \uparrow} (\tau) c^\dagger_{\vec{k} \uparrow} \rangle$ is the normal electron Green's function and $F(\vec{k},\tau) = - \langle T_\tau c_{\vec{k} \uparrow} (\tau) c_{-\vec{k} \downarrow} \rangle $ the anomalous (Gorkov) Green's function. Using Dyson's equation, the effect of interactions is captured by the self-energy, which is now also a matrix
\begin{equation}
	\hat{G}^{-1} (\vec{k}, i \omega_n) = 
		\hat{G}_0^{-1} (\vec{k}, i \omega_n)
		- \hat{\Sigma}(\vec{k}, i \omega_n),
	\label{Eq:Dyson}
\end{equation}
conveniently expressed in terms of fermionic Matsubara frequencies $\omega_n = (2n+1) \pi \kB T$. The bare Green's function is diagonal in spinor space,
\begin{equation}
	\hat{G}_0^{-1}(\vec{k},i\omega_n) = i\omega_n\hat{\tau}^0
	+ \left(\mu - \varepsilon_{\vec{k}}\right)\hat{\tau}^3,
\end{equation}
where $\hat{\tau}^3 = {\footnotesize\begin{pmatrix} 1&0\\0&-1 \end{pmatrix}}$ is the third Pauli matrix and $\hat{\tau}^0$ the identity, the chemical potential $\mu$ determines the electronic density $\ne$, and $\varepsilon_{\vec{k}}$ is the dispersion relation of free electrons,
\begin{equation}
	\varepsilon_{\vec{k}}=\frac{\hbar^2k^2}{2\me}.
\end{equation}
So far, we have mostly introduced some basic notation. The core of ME theory is a self-consistent summation of non-crossing diagrams from the perturbation theory in $V_{\mathrm{eff}}$. This provides us with the self-energy matrix $\hat{\Sigma}$ and, in particular, the off-diagonal component $\Sigma_{12}$ which is nonzero in the superconducting phase. The self-consistent summation of non-crossing diagrams ignores vertex corrections, which are of order $\mathcal{O}(\me/\mp) \sim 10^{-3}$ following Migdal's theorem. Given the large mass ratio between electrons and protons, the vertex corrections can be safely ignored.

This brings us to the Migdal--Eliashberg self-consistent equation for the self-energy, which reads \footnote{A Hartree term, not included in Eq.~\eqref{Eq:SelfEnergy}, gives for the jellium model a real constant, which can be absorbed in $\mu$.}
\begin{equation}
	\hat{\Sigma}(\vec{k},i\omega_n)
	= - \kB T \sum_{i \omega_m} \int \frac{d^3p}{(2\pi)^3}\,
		V_{\rm eff} (\vec{k} - \vec{p}, i\omega_n - i \omega_m)
		\hat{\tau}^3 \hat{G}(\vec{p},i\omega_m) \hat{\tau}^3 e^{i \omega_m 0^+},
	\label{Eq:SelfEnergy}
\end{equation}
where the effective interaction $V_{\mathrm{eff}}$ is given by Eq.~\eqref{Eq:EffectiveInteraction}, $\hat{\Sigma}$ and $\hat{G}$ are the self-energy and Green's function matrix respectively, and the factor $e^{i \omega_m 0^+}$ arises from imaginary-time ordering and ensures convergence of the summation over the fermionic Matsubara frequency $i \omega_m$. The electronic density is given by
\begin{equation}
	\ne =2\kB T\sum_{i\omega_n}\int\frac{d^3k}{(2\pi)^3}\,
	G(\vec{k},i\omega_n)e^{i\omega_n0^+}.
	\label{Eq:ElectronicDensity}
\end{equation}
Just below the temperature where the system passes from normal to superconducting, the Eliashberg equations for different angular momenta decouple. We will consider the $\ell=0$ ($s$-wave) case, as it is typically the channel giving the highest $\Tc$ for a phonon-mediated pairing interaction, especially a spherically-symmetric one like in the jellium model. In Appendix~\ref{Appendix:HigherAngularMomentum}, we discuss higher angular momentum pairing. The $s$-wave self-energy is spherically symmetric, i.e., $\hat{\Sigma}(\vec{k},i\omega_n) = \hat{\Sigma}(k,i\omega_n)$, where $k = |\vec{k}|$. Because $\hat{G}_0$ is spherically symmetric as well, so is $\hat{G}$ according to Eq.~\eqref{Eq:Dyson}. This circumstance allows us to perform the angle integration over wavevector $\vec{p}$ exactly in spherical coordinates,
\begin{equation}
	\int \frac{d^3p}{(2\pi)^3} \rightarrow \int_0^\infty \frac{dp\,p^2}{4\pi^2} \int_0^\pi d\theta\, \sin \theta,
\end{equation}
because the interaction only depends on $|\vec{k}-\vec{p}|^2 = k^2 + p^2 - 2 kp \cos \theta$. Introducing for convenience $\Epn=\hbar\wpn$ and the bosonic Matsubara frequency $\Omega_m = 2 m \pi \kB T$, we obtain
\begin{multline}
	V_{\mathrm{eff}}(k,p,i\Omega_m) =
	\int_0^\pi d\theta\, \sin \theta\, V_{\mathrm{eff}}(\vec{k}-\vec{p},i\Omega_m) \\
		= \frac{e^2}{2 \epsilon_0 k p } \frac{\Omega_m^2}{\Omega_m^2 + \Epn^2} 
		\ln \left[ \frac{(k+p)^2 (\Omega_m^2 + \Epn^2) + \kTF^2 \Omega_m^2}
		{(k-p)^2 (\Omega_m^2 + \Epn^2) + \kTF^2 \Omega_m^2} \right],
	\label{Eq:AngleIntegrated}
\end{multline}
which yields the following self-consistent equation,
\begin{equation}
	\hat{\Sigma}(k, i\omega_n) = - \kB T \sum_{i \omega_m} \int_0^\infty \frac{dp\,p^2}{4\pi^2}\,
	V_{\mathrm{eff}}(k,p,i\omega_n - i \omega_m)
	\hat{\tau}^3 \hat{G}(p,i\omega_m) \hat{\tau}^3 e^{i \omega_m 0^+}.
	\label{Eq:EliashbergEquation}
\end{equation}
Solving Eq.~\eqref{Eq:EliashbergEquation} requires integrating over one momentum degree of freedom and summing over one set of Matsubara frequencies, which we will now do numerically.

\section{Numerical implementation}
\label{Sec:NumericalImplementation}

The unique parameter of the jellium model is the electron density $\ne$. Calculating $\Tc$ versus $\ne$ requires one to determine $\mu$ by inverting the relation $\ne(\mu)$ entailed by Eq.~\eqref{Eq:ElectronicDensity}. The dominant renormalization of the chemical potential comes from the real part of the normal-state self-energy. To compensate this effect, we subtract $\mathrm{Re}\,\Sigma_{11}(\kF, i \pi \kB T)$ from $\Sigma_{11}(k, i\omega_n)$ and correspondingly for $\Sigma_{22}$. We then run the self-consistent search with $\mu=\eF=\hbar^2/(2\me)(9\pi^4\ne\kern-0.35em^2)^{1/3}$. At self-consistency, we recompute $\ne$ using Eq.~\eqref{Eq:ElectronicDensity}. We find that it differs from the initial $\ne$ by less than 3.5\% over the whole range of densities and temperatures studied, showing that the correction applied to the self-energy captures most of the chemical-potential renormalization. On the following graphs, we display the value of $\ne$ evaluated at self-consistency, unless specified otherwise.

We faced two difficulties while solving Eq.~\eqref{Eq:EliashbergEquation}. The first arises because $\hat{G}(p, i\omega_n)$ is extremely peaked around the Fermi wavevector $\kF$, despite the fact that $\hat{\Sigma}(p, i\omega_n)$ is a smooth function of $p$. Adaptive $p$-meshes did not solve the problem, giving greatly different self-consistent results depending on the mesh structure and density. We obtained stable results with monotonic convergence properties by evaluating exactly the rapidly-varying part of the integral. Specifically, if $p$ is replaced by $\kF$ at the first line of Eq.~\eqref{Eq:EliashbergEquation} and if $\hat{\Sigma}(p, i\omega_n)$ is replaced by $\hat{\Sigma}(\kF, i\omega_n)$ in the definition of $\hat{G}(p, i\omega_n)$ at the second line, the resulting $p$-integral is known exactly. By adding and subtracting this analytical result in Eq.~\eqref{Eq:EliashbergEquation}, the part that remains to be evaluated numerically is smoother around $p=\kF$, where it vanishes. After comparing various adaptive meshes for that remaining part, we reached the conclusion that a uniform mesh offers the best convergence properties. Details and convergence studies can be found in Appendix~\ref{App:numerics}.

\begin{figure}[b]
\includegraphics[width=0.7\columnwidth]{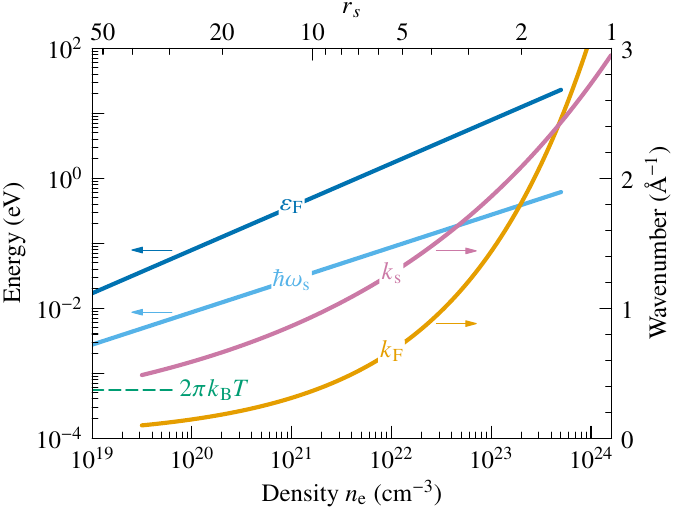}
\caption{\label{fig:energy-length}
Fermi energy $\eF$, plasma frequency of the charge compensating background $\wpn$ (left axis), Fermi wavevector $\kF$ and inverse Thomas-Fermi screening length $\kTF$ (right axis) as a function of the electronic density (bottom scale). The Wigner-Seitz radius, $r_s=(4\pi \ne/3)^{-1/3}/\aB$ with $\aB$ the Bohr radius, is indicated on the top scale. The dashed line indicates the energy difference between consecutive Matsubara frequencies for a temperature $T=1$~K.
}
\end{figure}

The second difficulty is that the number of frequencies $i\omega_m$ needed to achieve convergence increases as $1/T$ and gets enormous, because $\Tc$ is very low. At first sight, fast Fourier transforms would be the tool of choice for evaluating the $\omega_m$ sum in Eq.~\eqref{Eq:EliashbergEquation}. However, besides introducing boundary errors, this approach becomes impractical at very low $T$. We have followed a different route, motivated by the fact that the difference between consecutive $\omega_m$ values is much smaller than the typical energy scales $\eF$ and $\hbar\wpn$ over which $\Sigma_{11}$ and $\Sigma_{12}$ vary along the imaginary-frequency axis, respectively, i.e., $2\pi\kB T\ll\eF,\hbar\wpn$ in all our calculations (see Fig.~\ref{fig:energy-length}). In this regime, the discrete nature of Matsubara frequencies becomes irrelevant and the $\omega_m$ sum can be converted to an integral. In practice, we perform a discrete sum over the lowest frequencies and evaluate the rest as a continuous integral computed using quadratures, checking convergence with respect to the number of frequencies that are treated as discrete. To perform the quadrature, we construct functions of the \emph{continuous} variable $\omega_m$ that match the values computed at a subset of \emph{discrete} $\omega_m$. An advantage of this method is that the computation cost is the same at all temperatures. Another is that the frequency sum can be extended to infinity, avoiding an ultraviolet cutoff. Details are provided in Appendix~\ref{App:numerics}.

\begin{figure*}[tb]
\includegraphics[width=\textwidth]{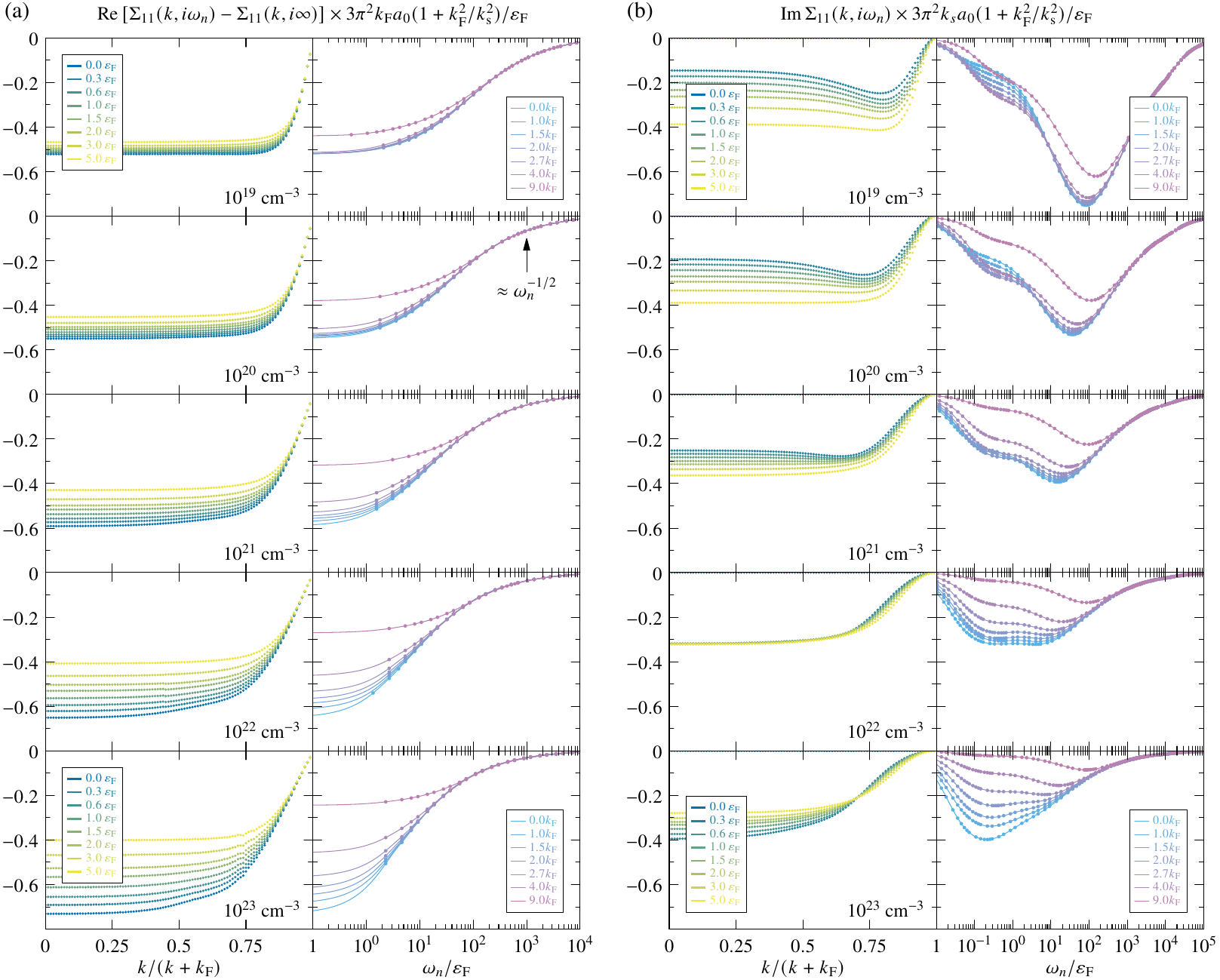}
\caption{\label{fig:Sigma11}
(a) Real and (b) imaginary parts of the self-consistent dynamical self-energy $\Sigma_{11}(k,i\omega_n)-\Sigma_{11}(k,i\infty)$, both as a function of reduced momentum $k/(k+\kF)$ and energy $\omega_n$ in units of $\eF$. The different rows correspond to different electronic densities (the values indicated are the densities at the start of the self-consistent search). The solid lines show the continuous compressions along the imaginary-frequency axis (see Appendix~\ref{App:numerics}). The dots show the compression anchor points, where the self-energy is reproduced exactly.
}
\end{figure*}

Note that the convergence factor $e^{i \omega_m 0^+}$ can be dropped for the calculation of the anomalous self-energy $\Sigma_{12}$, because $G_{12}(k, i\omega_m)$ behaves as $1/\omega_m^2$ at high frequency and the sum is therefore convergent. On the contrary, $G_{11}(k, i\omega_m)$ behaves as $1/\omega_m$ and the sum would diverge without that factor, because $V_{\mathrm{eff}}(k,p,i\Omega_m)$ approaches a constant value at high frequency. Here again, we add and subtract a term that we can sum exactly using
\begin{equation}
	\kB T \sum_{i \omega_m}G_0(p,i\omega_m)e^{i \omega_m 0^+} = \frac{1}{e^{(\varepsilon_p-\mu)/\kB T}+1},
\end{equation}
such that the remaining part evaluated numerically behaves as $1/\omega_m^2$ and doesn't need the convergence factor.

With the self-consistent solution of Eq.~\eqref{Eq:EliashbergEquation} at hand, we track the evolution of $\Sigma_{12}(k, i\omega_n)$ with increasing temperature $T$ until $\Tc$, where it vanishes. While $\Sigma_{12}(\kF, i\omega_0)$ gives the strength of pairing on the Fermi surface, the renormalization of the dispersion by the normal self-energy reduces the spectral gap. In several of the following figures, we display the ``gap'' $\Delta$ defined as
\begin{align}
	\label{Eq:Delta}
	\Delta(k,i\omega_n) &= \Sigma_{12}(k, i\omega_n)Z(k,i\omega_n) \\
	Z(k,i\omega_n) &= \left[1-\frac{\partial\mathrm{Im}\,\Sigma_{11}(k,i\omega_n)}{\partial\omega_n}\right]^{-1},
\end{align}
where $\partial$ is meant as a discrete derivative. Note that here, like in the Fermi-liquid literature \cite{Galitskii.1958, abrikosov2012methods}, $Z(k,i\omega_n)$ has the meaning of a quasiparticle residue, typically smaller than unity, in contrast to the traditional Eliashberg literature \cite{eliashberg1960interactions, parks1969superconductivity, marsiglio2020eliashberg}, where the same symbol represents a renormalization factor typically larger than unity (see Appendix~\ref{App:FactorZ} for a more in-depth discussion). To quantify the dynamical effect of $Z(k,i\omega_n)$, we will also report solutions of Eq.~\eqref{Eq:EliashbergEquation} obtained by setting the normal self-energy to zero.

\section{Numerical results}

We solve the Migdal--Eliashberg Eq.~\eqref{Eq:EliashbergEquation} for the jellium model with hydrogen ions ($Z_{\mathrm{I}} = 1$, $M = \mp$). Figure~\ref{fig:energy-length} compares the energy and length scales relevant for this system. We specifically look for superconductivity in a range of electronic densities from $\ne = 10^{19}$~cm$^{-3}$ to $10^{23}$~cm$^{-3}$. The first step is to find a self-consistent solution at $T = 1$~mK. We display the full momentum and energy dependencies of these solutions below. In a second step, we progressively increase the temperature towards $\Tc$. We find that, by and large, the only effect of increasing $T$ is a global rescaling of $\Sigma_{12}(k,i\omega_n)$. Therefore, we only display the temperature dependence of $\Sigma_{12}(\kF,i\omega_0)$ in the following, with $\omega_0=\pi\kB T$.

\subsection{Normal self-energy}

We first report the normal self-energy $\Sigma_{11}(k,i\omega_n)$ obtained after reaching self-consistency at $T=1$~mK. Figure~\ref{fig:Sigma11} shows the real and imaginary parts of the self-consistent $\Sigma_{11}(k,i\omega_n)$. The left panels show the solution versus $k/(k+\kF)$ at various values of $\omega_n$. The right panels show the solution versus $\omega_n$ at various values of $k$. The real static self-energy $\Sigma_{11}(k,i\infty)$ is subtracted, such that the plots only show the dynamical part of the self-energy. The real part has an approximately density-independent amplitude of order one when multiplied by $3\pi^2\kF\aB(1+\kF^2/\kTF^2)/\eF$, which corresponds to normalizing the self-energy by the quantity $(\ne/8\pi)V(\kF,i\infty)(\kTF/\kF)^2$. Note that, over the wide energy range of the right panels in Fig.~\ref{fig:Sigma11}(a), the self-energy has not yet entered the asymptotic regime, where it behaves as $1/\omega_n$. Instead, between $\sim 100\eF$ and $\sim 5000\eF$ there is a regime where $\mathrm{Re}\,\Sigma_{11}(k,i\omega_n)$ drops with a power law similar to $\omega_n^{-1/2}$. The transitions between the various regimes are more visible in the imaginary part. It turns out that the imaginary part of $\Sigma_{11}(k,i\omega_n)$ must be multiplied by $3\pi^2\kTF\aB(1+\kF^2/\kTF^2)/\eF$ in order to display with an approximately density-independent amplitude---i.e., a factor $\kTF/\kF$ with respect to the real part. The energy dependence of $\mathrm{Im}\,\Sigma_{11}(k,i\omega_n)$ displays several features that we briefly describe now.

All the structures of $\mathrm{Im}\,\Sigma_{11}(k,i\omega_n)$ in Fig.~\ref{fig:Sigma11}(b) are already present at lowest order, when Eq.~\eqref{Eq:EliashbergEquation} is evaluated with $\hat{\Sigma}(k,i\omega_m)=0$ on the right-hand side. In this case, the two relevant processes are the absorption of a phonon by a hole being filled by a Fermi-sea electron and the emission of a phonon by an electron above the Fermi sea. The first process is cut at the band bottom and explains the first structure of $\mathrm{Im}\,\Sigma_{11}(k,i\omega_n)$, which occurs at a largely momentum-independent energy $\omega_n\lesssim \eF$. The second process is unbounded in energy, because an electron injected at any energy and momentum can always relax to an eigenstate. Indeed, the kinetic constraint $\varepsilon-\hbar\omega_{\mathrm{ph}}(\vec{q})=\xi_{\vec{k}-\vec{q}}$ always admits a solution ($\xi_{\vec{k}}=\varepsilon_{\vec{k}}-\mu$). Therefore, unlike the absorption process, the emission process is not cut by kinematic constraints, but by the short-range Coulomb repulsion, which suppresses large-$q$ processes. For instance, if an electron is injected at $k=0$ and $\varepsilon\gg\eF$, it can relax to an eigenstate with energy $\varepsilon-\hbar\omega_{\mathrm{ph}}(\vec{q})\approx\varepsilon$---because $\hbar\omega_{\mathrm{ph}}(\vec{q})$ is bounded by $\hbar\wpn\ll\eF$---with emission of a phonon with $q\approx(2m\varepsilon/\hbar^2)^{1/2}$. The matrix element for this process involves $V(\vec{q},\varepsilon )$, which behaves as $1/(\kTF^2+q^2)$ at large $\varepsilon$. Hence this relaxation process typically behaves as $1/[(\kTF/\kF)^2+\varepsilon/\eF]$, which crosses over to $1/\varepsilon $ when $\varepsilon/\eF>(\kTF/\kF)^2$. For $\ne=10^{19}$~cm$^{-3}$, $(\kTF/\kF)^2\approx36$, while for $\ne=10^{23}$~cm$^{-3}$, $(\kTF/\kF)^2\approx1.7$. Indeed the high-energy structure in the right panels of Fig.~\ref{fig:Sigma11}(b) for $k=0$ moves down in energy from $\sim80\eF$ at $\ne=10^{19}$~cm$^{-3}$ to $\sim\eF$ at $\ne=10^{23}$~cm$^{-3}$. For $k\neq0$, the length of the matching vector $\vec{q}$ is reduced and consequently the crossover energy is increased. Very roughly, we have $q\approx(2m\varepsilon/\hbar^2)^{1/2}-k$, leading to a crossover energy near $\varepsilon/\eF=(\kTF/\kF+k/\kF)^2$. For $k=10\kF$, this energy varies from $256\eF$ at $\ne=10^{19}$~cm$^{-3}$ to $\sim128\eF$ at $\ne=10^{23}$~cm$^{-3}$, consistently with the data.

\begin{figure}[tb]
\includegraphics[width=0.5\columnwidth]{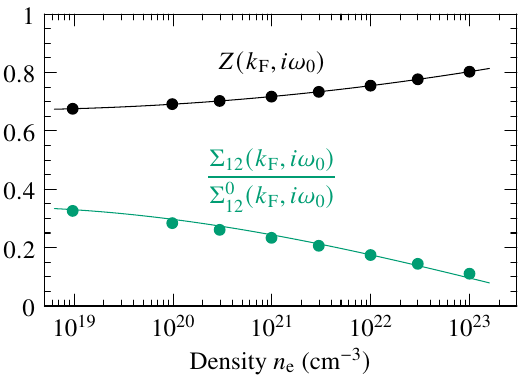}
\caption{\label{fig:residue}
Quasiparticle residue on the Fermi surface (black) and renormalization of the anomalous self-energy (green) versus self-consistent electron density. The solid lines are guides to the eye.
}
\end{figure}

The main renormalization effect of the normal self-energy is summarized in the quasiparticle residue evaluated on the Fermi surface and at the lowest Matsubara frequency, which is plotted in Fig.~\ref{fig:residue}. As the density grows, the kinetic energy progressively dominates the interaction energy and the residue approaches the noninteracting value $Z=1$. A quasiparticle residue $Z\sim0.65$--$0.8$ as found here corresponds to a coupling constant $1/Z-1\sim0.5$--$0.25$. 

The data displayed in Fig.~\ref{fig:Sigma11} underlines that converging the normal self-energy requires a set of Matsubara frequencies spanning an energy range several orders of magnitude larger than $\eF$. This seems to exclude any approach based on discrete frequencies. Yet, since the feedback of the high-frequency $\Sigma_{11}$ on the anomalous self-energy $\Sigma_{12}$ is suppressed by factors of order $\Sigma_{11}(k,i\omega_n)/\omega_n$, the self-consistent $\Sigma_{12}$ may be accurate even if $\Sigma_{11}$ is truncated or not converged at high frequency. However, it is mandatory to converge the high-frequency $\Sigma_{11}$ in order to enable the calculation of the density in Eq.~\eqref{Eq:ElectronicDensity}, because the convergence of the frequency sum in that expression is controlled by $-\mathrm{Re}\,\Sigma_{11}(k,i\omega_n)/\omega_n^2$ at high frequency. If $\Sigma_{11}$ is set by hand to zero, the convergence is faster and we find that $\ne$, as given by Eq.~\eqref{Eq:ElectronicDensity}, is equal to the starting density within our numerical accuracy---i.e., we find no noticeable renormalization of the chemical potential by the opening of the superconducting gap. With $\Sigma_{11}$ included, Eq.~\eqref{Eq:ElectronicDensity} gives a value very slightly smaller than the starting density, by a factor ranging from $0.965$ at $10^{19}$~cm$^{-3}$ to $0.998$ at $10^{23}$~cm$^{-3}$.

\subsection{Anomalous self-energy}

\begin{figure}[tb]
\includegraphics[width=0.7\textwidth]{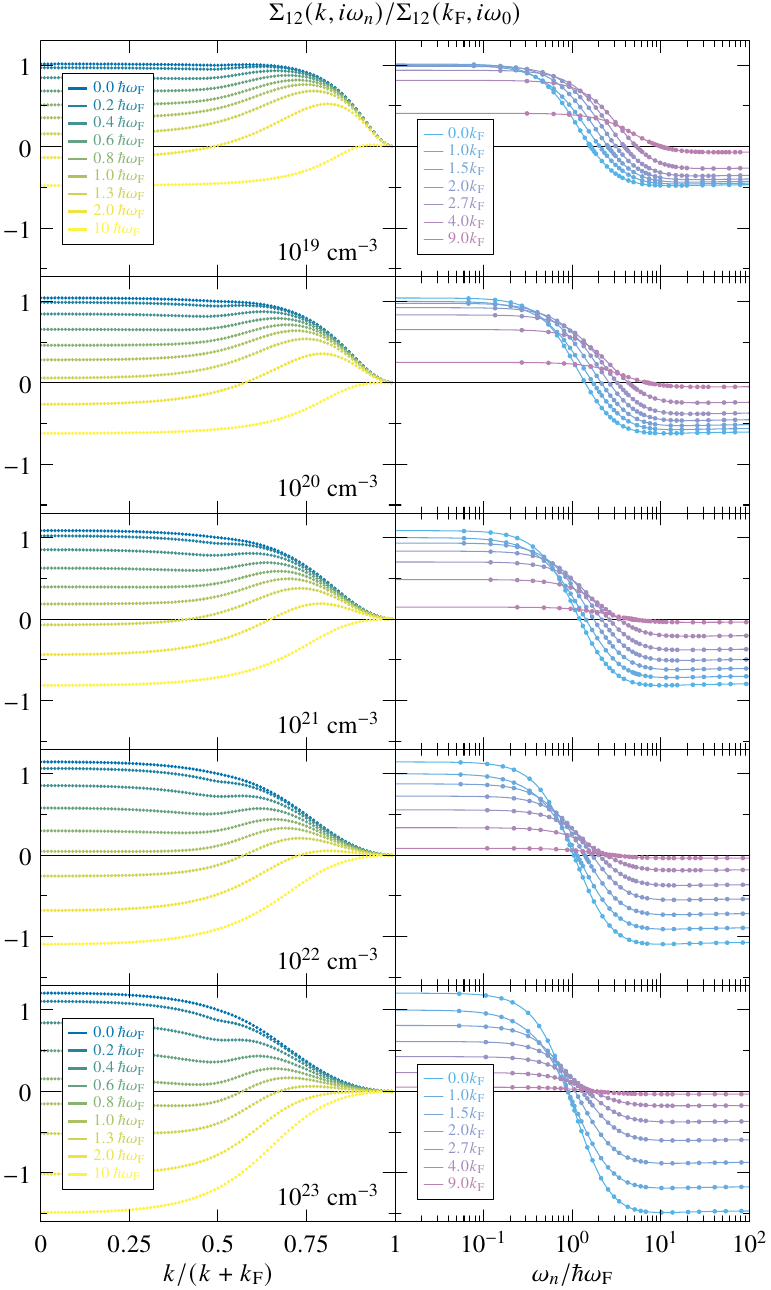}
\caption{\label{fig:Sigma12}
Self-consistent anomalous self-energy $\Sigma_{12}(k,i\omega_n)$ at $T=1$~mK for the (starting) values of $\ne$ indicated in the left panels. As in Fig.~\ref{fig:Sigma11}, we show both the dependence on the reduced momentum $k/(k+\kF)$ and energy $\omega_n$, however in units of $\hbar\omega_{\mathrm{F}}=\hbar\omega_{\mathrm{ph}}(\kF)$. The solid lines show the continuous compressions along the imaginary-frequency axis. The dots show the compression anchor points, where the anomalous self-energy is reproduced exactly.
}
\end{figure}

We can now turn to the properties of the superconducting phase. Figure~\ref{fig:Sigma12} shows the self-consistent solution $\Sigma_{12}(k,i\omega_n)$, normalized by $\Sigma_{12}(\kF,i\omega_0)$, at $T=1$~mK and for eight different densities. Here we measure the energies in units of $\hbar\omega_{\mathrm{F}}=\hbar\omega_{\mathrm{ph}}(\kF)$. At low energy, the gap is positive and decreases progressively to zero as $k$ grows past $\kF$ (blue lines). This behavior is smoother at higher density. At any wavevector, the gap turns from positive to negative as the energy increases, changing sign typically at $\hbar\omega_{\mathrm{F}}$ (right panels). At low density, the gap on the Fermi surface (lightblue lines) is smaller in magnitude at infinite frequency as compared to zero frequency. The situation is opposite at high density.

\begin{figure}[tb]
\includegraphics[width=0.5\columnwidth]{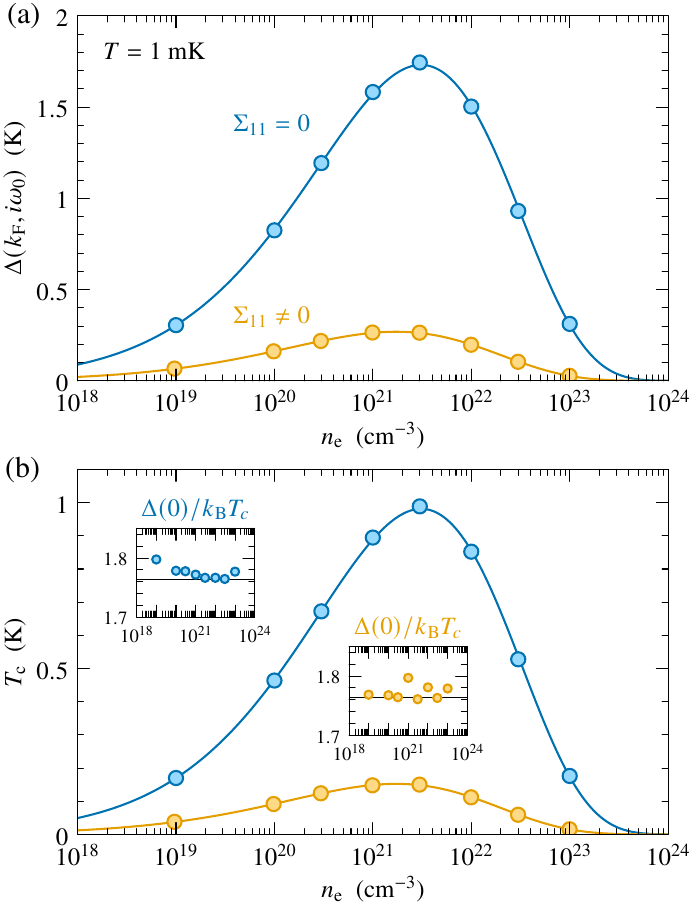}
\caption{\label{fig:Delta-Tc-n}
(a) Superconducting gap at 1~mK on the Fermi surface for the lowest Matsubara frequency $\omega_0 = \pi\kB T$, expressed in Kelvin, and (b) critical temperature, in the cases without (blue) and with (orange) the normal self-energy $\Sigma_{11}(k,i\omega_n)$. The solid lines are guides to the eye. Insets: ratio $\Delta(0)/(\kB\Tc)$ compared with the BCS value (horizontal line).
}
\end{figure}

We also calculated the self-consistent solution $\Sigma_{12}^0$ for the case where the normal self-energy $\Sigma_{11}$ is discarded. This solution gives plots almost indistinguishable from those in Fig.~\ref{fig:Sigma12}, showing that, to a very good approximation, the normal self-energy just provides a rigid renormalization of the anomalous one, i.e., $\Sigma_{12}\approx\Sigma_{12}^0R$. The renormalization factor $R$ is different from the quasiparticle residue $Z$, because the latter renormalizes the energy in the nonlinear self-consistent equation giving the former. As a result, $R$ is a nonlinear function of $Z$. This is illustrated in Fig.~\ref{fig:residue}. Over the range of densities of the plot, we have $R\approx 1/Z-1.15$ (green line). Since $T_c$ is proportional to $\Delta$ as defined in Eq.~\eqref{Eq:Delta}, the expected renormalization of the critical temperature by the normal self-energy is $RZ\approx1-1.15Z$ for the densities considered.

From the anomalous self-energy $\Sigma_{12}(\kF,i\omega_0)$ used to normalize the data in Fig.~\ref{fig:Sigma12} and the normal self-energy $\Sigma_{11}(\kF,i\omega_n)$ displayed in Fig.~\ref{fig:Sigma11}, we deduce the spectral gap $\Delta(\kF,i\omega_0)$ according to Eq.~\eqref{Eq:Delta}, which is plotted versus density in Fig.~\ref{fig:Delta-Tc-n}(a). For comparison, we also show the case where $\Sigma_{11}$ is neglected. In the latter case, the largest gap is found at $\ne\approx3\times10^{21}$~cm$^{-3}$ while in the former the maximum occurs close to $1.8\times10^{21}$~cm$^{-3}$. The dome shape results from a competition between $\eF$ (or $\wpn$) increasing with increasing $\ne$ and the coupling strength being reduced due to increasing screening.

\begin{figure}[tb]
\includegraphics[width=0.5\columnwidth]{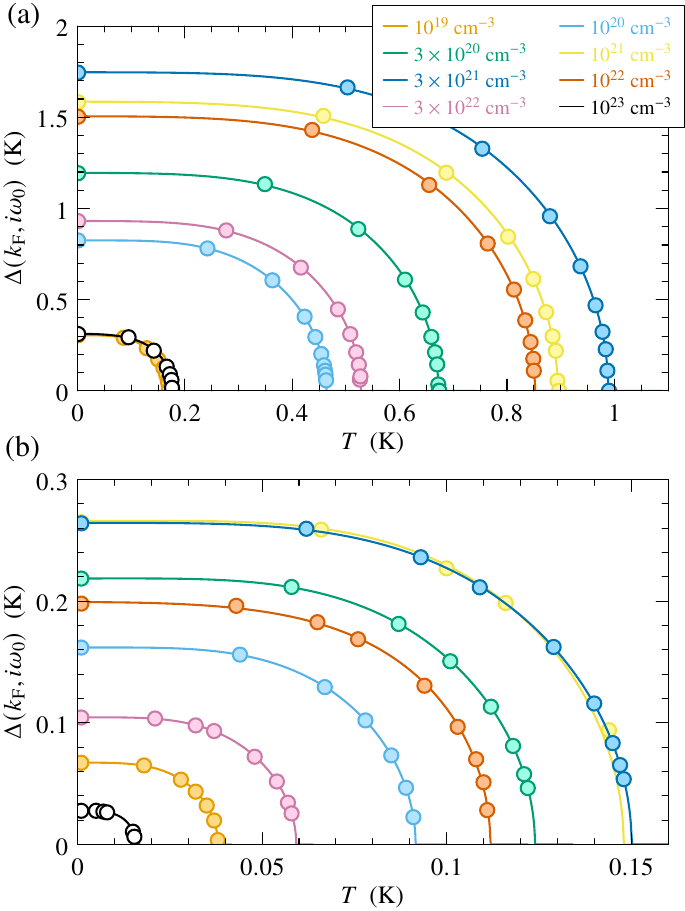}
\caption{\label{fig:Delta-T}
Self-consistent gap function on the Fermi surface versus temperature (dots) and fitted temperature dependence [lines, following Eq.~\eqref{Eq:DeltaFitTDependence}] in the absence (a) / presence (b) of $\Sigma_{11}(k,i\omega_n)$.
}
\end{figure}

The temperature dependence of the self-consistent solution can be largely summarized in the temperature dependence of $\Delta(\kF,i\omega_0)$. Indeed, the full solutions at $T\lesssim\Tc$ yield figures in all ways similar to Fig.~\ref{fig:Sigma12}. Likewise, the self-consistent $\Sigma_{11}(k,i\omega_n)$ is essentially independent of $T$ up to $\Tc$. Figures~\ref{fig:Delta-T}(a) and \ref{fig:Delta-T}(b) show how the gap closes with increasing $T$ when the normal self-energy is discarded, respectively included. A functional form 
\begin{equation}
	\Delta(T)=\Delta(0)\sqrt{1-(T/\Tc)^2}\left[1+a(T/\Tc)^2e^{-b(T/\Tc)^2}\right]
	\label{Eq:DeltaFitTDependence}
\end{equation}
is fit to each set of data in order to extract $\Tc$ and the ratio $\Delta(0)/\Tc$. The parameters $a$ and $b$ are typically $0.5$ and $1$, with variations from one density to the next. The extracted $\Tc$ are reported in Fig.~\ref{fig:Delta-Tc-n}(b), with a maximum $\Tc^{\max}\approx1$~K at $\ne\approx3\times10^{21}$~cm$^{-3}$ without normal self-energy and a maximum $\Tc^{\max}\approx150$~mK at $1.8\times10^{21}$~cm$^{-3}$ with the self-energy. The gap closes in the mean-field fashion in both cases and we find that the ratio $\Delta(0)/(\kB\Tc)$ is consistent with the BCS value $\pi/e^\gamma=1.764$ [insets of Fig.~\ref{fig:Delta-Tc-n}(b)].

\section{Comparison to approximation schemes}

\begin{figure}[tp]
\includegraphics[width=0.5\columnwidth]{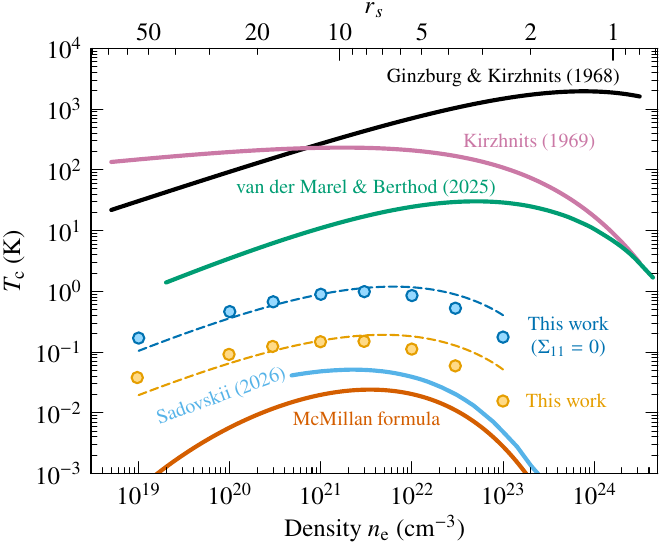}
\caption{\label{fig:summary}
$\Tc$ as a function of density (bottom scale) calculated with different approximations: On-energy-shell (black \cite{ginzburg1968superconductivity}, purple \cite{kirzhnits1969superconductivity}, green \cite{marel2024superconductivity}); on-momentum-shell (dashed lines, this work); the solution of the Kirzhnits--Maksimov--Khomskii \cite{kirzhnits1973the} integral equation (lightblue \cite{sadovskii2026}); McMillan formula (orange \cite{marel2024superconductivity}); full momentum-dependent Eliashberg (circles, this work). The Wigner-Seitz radius, $r_s=(4\pi \ne/3)^{-1/3}/\aB$ is indicated on the top scale.
}
\end{figure}

With inclusion of retardation and normal self-energy, we find that the hydrogen jellium model only reproduces a meager maximum $\Tc$ of 150~mK. This is in stark contrast with earlier estimates of $\Tc$ in excess of 1000~K. In Fig.~\ref{fig:summary}, estimates for $\Tc$ are summarized, each obtained with a different approach: the fully momentum- and frequency-dependent Eliashberg equations of this work, three different implementations of the ``on-energy-shell'' substitution $\hbar\omega=\varepsilon_p-\varepsilon_k$ \cite{ginzburg1968superconductivity, kirzhnits1969superconductivity, marel2024superconductivity}, the Eliashberg equations in the $k=p=\kF$ approximation (see Sec.~\ref{Sec:OnMomentumShell}), a recent Eliashberg calculation with a simplified angle-averaged interaction \cite{sadovskii2026}, and the $\Tc$'s from the McMillan formula \cite{mcmillan1968transition} (see Sec.~\ref{Sec:McMillan}).

The ``on-energy-shell'' approximation $\hbar\omega=\varepsilon_p-\varepsilon_k$ used by Ginzburg--Kirzhnits \cite{ginzburg1968superconductivity}, Kirzhnits \cite{kirzhnits1969superconductivity} and two of us \cite{marel2024superconductivity} does not include a summation over (Matsubara) frequencies. The absence of explicit retardation in this approach seems to vastly overestimate $\Tc$. We found that we can get much more realistic estimates of $\Tc$ using two different approximations: an ``on-momentum-shell'' approximation, and by using the McMillan formula.

\subsection{On-momentum-shell approximation}
\label{Sec:OnMomentumShell}

As we discussed in Sec.~\ref{Sec:NumericalImplementation}, the Green's functions are highly peaked near the Fermi momentum. At the same time, the self-energies are varying slowly around the Fermi surface. This suggests that we can approximate $\hat{\Sigma}(k,i\omega_n)$ by $\hat{\Sigma}(\kF,i\omega_n)$ and ignore the full momentum-dependence of the interaction by replacing $p^2V_{\mathrm{eff}}(\kF,p,i\Omega_m)$ by $\kF^2V_{\mathrm{eff}}(\kF,\kF,i\Omega_m)$ in Eq.~\eqref{Eq:EliashbergEquation}. Note that $V_{\mathrm{eff}}(\kF,p,i\Omega_m)$ drops as $1/p^2$ as large $p$, which justifies using $\kF^2$ instead of $p^2$ for the volume integration. With this replacement, the self-consistent equations become
\begin{align}
	\hat{G}^{-1}(k,i\omega_n) &= \hat{G}_0^{-1}(k,i\omega_n)-\hat{\Sigma}(i\omega_n)\\
	\hat{\Sigma}(i\omega_n) &=-\kB T\sum_{i\omega_m}\int_0^\infty\frac{dp\,\kF^2}{4\pi^2}\,
	V_{\mathrm{eff}}(\kF,\kF,i\omega_n-i\omega_m)
	\hat{\tau}^3\hat{G}(p,i\omega_m)\hat{\tau}^3e^{i \omega_m 0^+}.
\end{align}
The $p$-integral is convergent and exactly known (see Appendix~\ref{App:numerics}) and the sum over the quasi-continuous frequencies can again be evaluated using quadratures. This ``on-momentum-shell'' approximation leads to results for $\Tc$ that are very close to the numerically exact solutions, as can be seen in Fig.~\ref{fig:summary}.

\subsection{McMillan approximation}
\label{Sec:McMillan}

A traditional approximation of the $\Tc$ that goes beyond the BCS equation is the McMillan formula \cite{mcmillan1968transition}
\begin{equation}
	\kB\Tc = \frac{\hbar\wpn}{1.45}\exp\left\{-\frac{1.04\,(1+\lambda)}
	{\lambda-\bar{\mu}\frac{1+0.62\lambda}{1+\bar{\mu}\ln(\eF/\hbar\wpn)}}\right\},
\end{equation}
where, for the jellium model, the phonon-mediated coupling $\lambda$ and the average Coulomb repulsion $\bar{\mu}$ are equal, $\lambda=\bar{\mu}=\ln(1+\pi\kF\aB)/(2\pi\kF\aB)$ \cite{marel2024superconductivity}. The exact cancellation between $\bar{\mu}$ and $\lambda$ suggests indeed that the corresponding $\Tc$ will naturally be very low. Note, however, that this cancellation doesn't occur when the charge compensating background has a finite compressibility \cite{marel2025thoughts} (see Appendix~\ref{Appendix:HigherAngularMomentum}).

Since the McMillan formula was obtained by fitting a variational expression to numerical solutions of the Eliashberg equations \cite{mcmillan1968transition}, one may in general expect the $\Tc$ following from the McMillan and the Eliashberg equations to be relatively close to each other. That in the present case of the jellium model the difference is rather important reflects in part that a situation where $\bar{\mu}=\lambda$ takes an extremal position relative to the cases considered by McMillan.

\section{Conclusions and Outlook}

We found that the superconducting critical temperature $\Tc$ derived from the jellium model at a given density depends dramatically on the approximation used, ranging from thousands of Kelvin to tens of milliKelvin. We have solved the full energy- and momentum-dependent Eliashberg equations and obtained a maximum $\Tc$ of $150$~mK at a density $\sim 1.8\times 10^{21}$~cm$^{-3}$. The computation is surprisingly heavy given the simplicity of the model and reveals that insufficient numerical convergence can also lead to wide variations in the calculated $\Tc$. Our results question the standard method for calculating $\Tc$ of real materials, where often the momentum dependence of the Eliashberg equations and the normal self-energy are ignored.

A limitation of the jellium model---and the related Bardeen--Pines model---is that the electronic screening is described using the Thomas--Fermi approximation. Since this approximation captures neither Friedel oscillations nor the Kohn--Luttinger pairing mechanism \cite{kohn1965new} (see Appendix~\ref{Appendix:HigherAngularMomentum}), this implies that, even if a jellium could be realized experimentally, there is still reason to believe that important aspects of the physics are left out, in particular the Friedel oscillations. This aspect is remedied in the random phase approximation (RPA) for the description of the charge screening. To capture spin-fluctuations mediated processes requires even further additional diagrams, however.

While the jellium model is by construction a drosophila, a test-subject for more complicated realistic models, it does contain the core ingredients thought necessary for superconductivity. Given the numerical challenges in solving the full momentum- and frequency-dependence of the jellium model, these numerical challenges will be even bigger in realistic models. Nevertheless, compared to BCS, momentum- or frequency-independent approximation schemes, an exact solution for the Eliashberg equations is necessary---after all, we show that even in the simple jellium model getting a quantitative estimate of $\Tc$ is difficult, and getting a \emph{high} $\Tc$ impossible.

\section{Acknowledgements}
We thank Peter Hirschfeld and Frank Marsiglio for inspiring discussions.
L.R.\ acknowledges the Swiss National Science Foundation (SNSF) via Starting Grant TMSGI2 211296.

\appendix

\section{Numerical tricks for a reliable evaluation Eq.~\eqref{Eq:EliashbergEquation}}
\label{App:numerics}

The numerical implementation of Eq.~\eqref{Eq:EliashbergEquation} needs careful attention due to Green's functions strongly peaked at the Fermi momentum and very dense Matsubara frequencies at low $T$. We first describe our strategy for the momentum integration, then for the frequency summation.

\subsection{Momentum integration}

To perform the rapidly-varying part of the momentum integral exactly, we introduce the ``Fermi-surface'' Green's function
\begin{equation}
	\hat{G}^{-1}_{\mathrm{F}}(p,i\omega_n) = \hat{G}_0^{-1}(p,i\omega_n)-\hat{\Sigma}(\kF,i\omega_n),
\end{equation}
whose $p$-integral is known:
\begin{align*}
	\int_0^{\infty}dp\,G_{\mathrm{F},11}(p,i\omega_m) &= -\sqrt{\frac{2\me}{\hbar^2}}\frac{i\pi}{4\gamma}
	\left[\frac{\omega_m-\Sigma_{11}''+\gamma}{\sqrt{\mu-\Sigma_{11}'+i\gamma}}
	-\frac{\omega_m-\Sigma_{11}''-\gamma}{\sqrt{\mu-\Sigma_{11}'-i\gamma}}\right]\\
	\int_0^{\infty}dp\,G_{\mathrm{F},12}(p,i\omega_m) &= -\sqrt{\frac{2\me}{\hbar^2}}\frac{\pi}{2}
	\frac{\Sigma_{12}}{\gamma}\mathrm{Re}\,\frac{1}
	{\sqrt{\mu-\Sigma_{11}'+i\gamma}},
\end{align*}
with the short-hand notations
\begin{align*}
	\Sigma_{11}(\kF,i\omega_m)&=\Sigma_{11}'+i\Sigma_{11}''\\
	\Sigma_{12}(\kF,i\omega_m)&=\Sigma_{12}\\
	\gamma & = \sqrt{(\omega_m-\Sigma_{11}'')^2+\Sigma_{12}^2}.
\end{align*}
After addition and subtraction of this expression, the numerical momentum integral in Eq.~\eqref{Eq:EliashbergEquation} involves 
\begin{equation}
	p^2V_{\mathrm{eff}}(k,p,i\omega_n-i \omega_m)\hat{G}(p,i\omega_m)-
	\kF^2V_{\mathrm{eff}}(k,\kF,i\omega_n-i \omega_m)\hat{G}_\mathrm{F}(p,i\omega_m),
\end{equation}
which vanishes at $p=\kF$. We recast the integral to a unit interval by means of the dimensionless variable $\kappa=p/(p+\kF)$. The self-consistent solution $\hat{\Sigma}(\kappa,i\omega_n)$ is defined on a discrete mesh of $M$ values of $\kappa$. We have compared meshes with various distributions of points, all of them having a point anchored at $\kF$ ($\kappa=1/2$). In practice, we distribute the mesh points uniformly along the arc-length of the function $|G_{12}(\kappa,i\omega_0)/G_{12}(1/2,i\omega_0)|^{\eta}$. The mesh tracks the $\kappa$-dependence of $G_{12}$ at the lowest Matsubara frequency with $\eta=1$, while with $\eta=0$ it is uniform in $\kappa$ space, which means having half the points between $k=0$ and $k=\kF$, and the other half between $\kF$ and $\infty$.

\begin{figure}[tb]
\includegraphics[width=0.7\columnwidth]{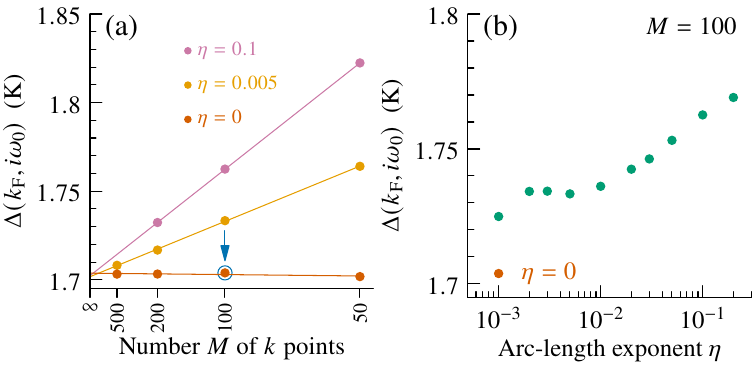}
\caption{\label{fig:M-eta}
Convergence with respect to the $k$-point mesh parameters. (a) Scaling with the number of $k$ points and extrapolation to a dense mesh for three values of $\eta$. The blue arrow indicates our default parameters $M=100$ and $\eta=0$. (b) Evolution with the arc-length exponent $\eta$ for $M=100$. The other numerical parameters are $N=10$ and $c_{\mathrm{tol}}=10^{-3}$ (see below). The physical parameters are $\ne=5\times10^{21}$~cm$^{-3}$ and $T=0.001$~K. The normal self-energy is set to zero.
}
\end{figure}

A typical convergence study with respect to the properties of the $k$-point mesh is displayed in Fig.~\ref{fig:M-eta}(a). Here we plot the self-consistent ``Fermi-surface gap'' $\Delta(\kF,i\omega_0)$ expressed in Kelvin, calculated at $T=1$~mK for $\ne=5\times10^{21}$~cm$^{-3}$ and ignoring the normal-state self-energy $\Sigma_{11}$ ($\Delta=\Sigma_{12}$ in that case). Similar results are found for the full model with self-consistent $\Sigma_{11}$ and for other electronic densities and temperatures. The scaling with $M$ allows a reliable extrapolation towards the infinitely dense mesh, to a value that is independent of $\eta$ within less than 0.1\%. Interestingly, the uniform mesh turns out to provide the best convergence properties with respect to $M$: The gap obtained with $M=100$ and $\eta=0$ agrees with the average extrapolated value within 0.08\%. The dependence on the exponent $\eta$ is shown in Fig.~\ref{fig:M-eta}(b). Exponents closer to 1---i.e., meshes more densely packed around $\kF$---tend to give larger self-consistent gap values, like less dense meshes at fixed $\eta$ do in Fig.~\ref{fig:M-eta}(a). These trends indicate that, after analytical integration of the region near $\kF$, the remaining information is not localized near $\kF$ and is therefore missed by meshes that are strongly focussing there ($\eta\sim1$), or that are focussing less ($\eta=0.005$), but not dense enough.

\subsection{Frequency summation}

We approximate the sums over an infinite set of discrete Matsubara frequencies $\omega_m$ by continuous integrals evaluated using quadratures, which is justified at the low temperatures of interest for the jellium model. To test the accuracy of this approximation, we start the integral beyond the $N+1$ lowest frequencies that are treated as discrete, according to
\begin{equation}
	\sum_{m=0}^{\infty}(\cdots)\approx\sum_{m=0}^N(\cdots)+\int_{N+1/2}^{\infty}dm\,(\cdots).
\end{equation}
As the integral needs the property $(\cdots)$ at arbitrary values of $\omega_m$, we build continuous functions $\Sigma_{11}^k(m)$ and $\Sigma_{12}^k(m)$ that are requested to match the calculated $\Sigma_{11}(k,i\omega_m)$ and $\Sigma_{12}(k,i\omega_m)$ exactly at recursively chosen values of $m$, while at the other values of $m$ a deviation by a relative tolerance $c_{\mathrm{tol}}$ is permitted. We call this process \emph{compression}, as the information available at the infinite set of Matsubara frequencies is compressed into a piecewise-continuous function containing a number of pieces as small as possible, such as to meet the compression criterion $c_{\mathrm{tol}}$. Illustrations of this compression, which is done independently for $\Sigma_{11}$ and $\Sigma_{12}$ at each point of the momentum mesh, can be seen in Figs.~\ref{fig:Sigma11} and \ref{fig:Sigma12}.

\begin{figure}[b]
\includegraphics[width=0.7\columnwidth]{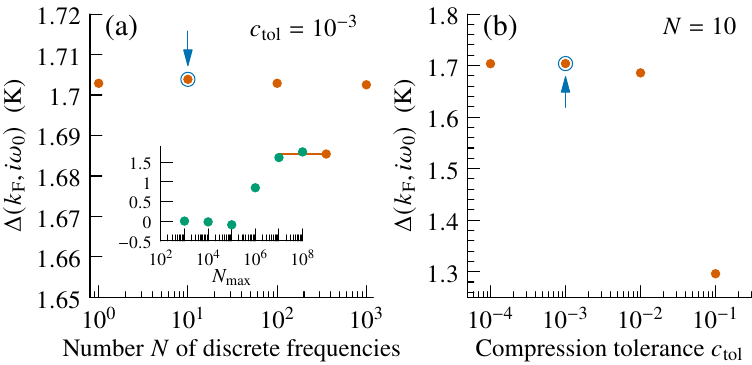}
\caption{\label{fig:N-ctol}
Convergence with respect to the frequency grid parameters. (a) Immediate convergence with the number $N$ of discrete frequencies. Inset. Slow convergence of the truncated discrete sum (see text). (b) Evolution with the compression tolerance $c_{\mathrm{tol}}$. The blue arrows indicate our default parameters $N=10$ and $c_{\mathrm{tol}}=10^{-3}$. The other numerical parameters are $M=100$ and $\eta=0$. The physical parameters are $\ne=5\times10^{21}$~cm$^{-3}$ and $T=0.001$~K. The normal self-energy is set to zero.
}
\end{figure}

Figure~\ref{fig:N-ctol}(a) shows that the discrete nature of Matsubara frequencies plays no significant role at the temperatures of interest, such that the value of $N$ is actually immaterial. We use $N=10$ in our calculations. For comparison, we also display as an inset in Fig.~\ref{fig:N-ctol}(a) the result obtained with just $\sum_{m=0}^{N_{\max}}(\cdots)$ versus $N_{\max}$. The result of this truncated sum has the wrong sign for $N_{\max}\lesssim10^5$ and only starts approaching convergence for $N_{\max}\gtrsim10^6$. At convergence, $N_{\max}\sim10^8$, the largest frequency $\omega_{N_{\max}}\sim 50\eF$ illustrates the impractical need to include millions of frequencies when treating them as discrete. Figure~\ref{fig:N-ctol}(b) shows that a compression tolerance of $10^{-3}$ yields a result well within $0.02\%$ of the value converged with respect to $c_{\mathrm{tol}}$.

$M$ and $c_{\mathrm{tol}}$ are the numerical parameters that affect the computation time, while $\eta$ and $N$ don't. Using $M=100$ and $c_{\mathrm{tol}}=10^{-3}$, we are confident that we stay within less than $1\%$ of the exact result. In line with our expected accuracy $\lesssim1\%$, we stop the self-consistent search when the relative variation of $\Sigma_{12}(k,i\omega_0)$ from one iteration to the next becomes less than $10^{-4}$, which means that the first four digits of our solutions are converged.

\section{\boldmath The factor $Z$}
\label{App:FactorZ}

In traditional Eliashberg literature \cite{eliashberg1960interactions, parks1969superconductivity, marsiglio2020eliashberg}, the self-energy matrix is split into three components $Z$, $\chi$, and $\phi$ following the expression
\begin{equation}
	\hat{\Sigma}(\vec{k},i \omega_n) = i \omega_n [1 - Z(\vec{k},i \omega_n)] \hat{\tau}^0 
	+ \phi(\vec{k},i \omega_n) \hat{\tau}^1 + \chi(\vec{k},i \omega_n) \hat{\tau}^3,
\end{equation}
such that the anomalous self-energy equals $\Sigma_{12}(\vec{k}, i \omega_n) = \phi(\vec{k}, i\omega_n)$. The anomalous self-energy is real and even in frequency, $\phi(\vec{k}, i \omega_n) = \phi(\vec{k}, - i \omega_n)$. Because the normal self-energy satisfies $\Sigma_{11}^*(\vec{k}, i \omega_n) = \Sigma_{11} (\vec{k}, - i \omega_n)$, it follows that $Z$ and $\chi$ are real even functions of frequency.

In this language, the anomalous Gorkov Green's function becomes
\begin{equation}
	F(\vec{k}, i\omega_n) = \frac{- \phi_{\vec{k},n}}
	{Z_{\vec{k},n}^2 \omega_n^2 + (\xi_{\vec{k}} + \chi_{\vec{k},n})^2 + \phi_{\vec{k},n}^2}
\end{equation}
where for notational convenience we write subscript $[\cdots]_{\vec{k},n}$ for $[\cdots](\vec{k}, i \omega_n)$. The poles of the corresponding real-frequency Green's function describe the Bogoliubov quasiparticles, which are obtained by solving 
\begin{equation}
	E^2 = \frac{(\xi_{\vec{k}} + \chi_{\vec{k},n})^2}{Z_{\vec{k},n}^2}
	+ \frac{\phi_{\vec{k},n}^2}{Z_{\vec{k},n}^2}.
\end{equation}
Recall that in BCS theory, the Bogoliubov quasiparticle dispersion is given by $E_{\vec{k}}^2 = \xi_{\vec{k}}^2 + \Delta_{\vec{k}}^2$. It is therefore natural to define the gap function as
\begin{equation}
	\Delta_{\vec{k},n} = \phi_{\vec{k},n} / Z_{\vec{k},n}.
\end{equation}

However, in the original Fermi-liquid literature \cite{Galitskii.1958, abrikosov2012methods}, the factor $Z$ refers to the quasiparticle weight and is defined through
\begin{equation}
	Z_{\vec{k},n} = \left[1 - \frac{\partial\, \mathrm{Im}\, \Sigma_{11}(\vec{k},i \omega_n)}
	{\partial \omega_n} \right]^{-1}.
\end{equation}
In the limit of small $\omega_n$, the Fermi-liquid $Z_{\vec{k},n}$ is the inverse of the Eliashberg $Z_{\vec{k},n}$! This can cause quite some confusion.

\section{Higher angular momentum order parameter}
\label{Appendix:HigherAngularMomentum}

\begin{figure}[tb]
\includegraphics[width=0.7\columnwidth]{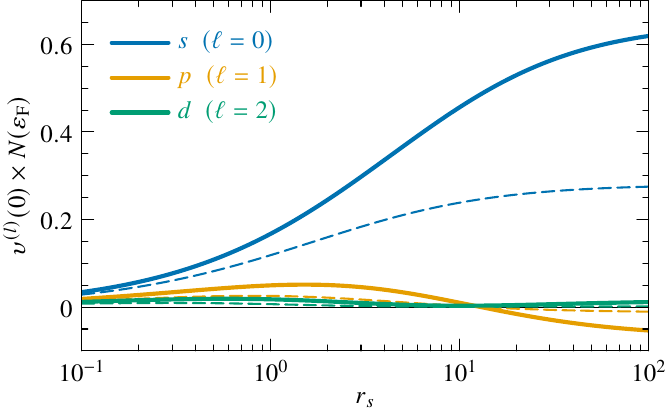}
\caption{\label{fig:angular-momenta}
Static electronically screened Coulomb interaction in the $s$-wave, $p$-wave, and $d$-wave channels as a function of $r_s$. The dashed lines show the interaction including screening by the phonons for $s=0$ and $4 k_F \cpn^2=\pi \aB \wpn^2 $.
}
\end{figure}

The interaction in Eq.~\eqref{Eq:AngleIntegrated} can be decomposed in angular-momentum channels according to
\begin{equation}\label{eq:interaction}
	V_{\mathrm{eff}}(\vec{k}-\vec{p},i\Omega_m)=4\pi\sum_{\ell m}
	Y_{\ell}^m(\hat{\vec{k}})^*\upsilon_{kp}^{(\ell)}(i\Omega_m)Y_{\ell}^m(\hat{\vec{p}}),
\end{equation}
where the $\ell$-component of the interaction is given by
\begin{equation}
	\upsilon_{kp}^{(\ell)}(i\Omega_m)=\frac{1}{2kp}\int_{|k-p|}^{|k+p|}\kern-0.5em
	dq\,qV_{\mathrm{eff}}(q,i\Omega_m)
	P_{\ell}\left(\textstyle\frac{k^2+p^2-q^2}{2kp}\right)
	\label{eq:interaction_ell}
\end{equation}
and $P_{\ell}(x)$ are Legendre polynomials. Just below the temperature where the system passes from normal to superconducting, the Eliashberg equations for different angular momenta decouple. For electrons at the Fermi surface, the interaction in the $\ell$-wave pairing channel including the phonon contributions is given by $\upsilon_{\kF\kF}^{(\ell)}(i\Omega_m)\equiv\upsilon^{(\ell)}(i\Omega_m)$:
\begin{equation}
	\upsilon^{(\ell)}(i\Omega_m)=\frac{1}{2\kF^2}\int_0^{2\kF}\kern-0.5em
	dq\,q\frac{e^2}{\epsilon_0q^2\epsilon(q,i\Omega_m)}
	P_{\ell}\left(\textstyle1-\frac{q^2}{2\kF^2}\right).
\end{equation}
Within the random phase approximation (RPA) for the jellium model, the dielectric function is
\begin{equation}
	\epsilon(q,i\Omega_m)=1-\frac{e^2}{\epsilon_0q^2}\Pi_0(q,i\Omega_m)
	+\frac{(\hbar\wpn)^2}{\Omega_m^2+\frac{(\hbar\cpn q)^2}{1 + (qs)^2}},
\end{equation}
where $\Pi_0$ is the noninteracting electron polarizability and $s$ is the lower length scale of sound-like dispersion \cite{marel2025thoughts}. The polarizability is described by the Lindhard function, which for $T=0$ and defining $x=q/(2\kF)$, $\varepsilon_q=\hbar^2q^2/(2\me)$, has the form \cite{mahan2000}
\begin{align}\label{eq:Pi0}
	\nonumber
	&\hspace{-1em}-\frac{e^2}{\epsilon_0}\Pi_0(q,i\Omega_m)=\kTF^2\left\{\frac{1}{2}\,+\right.\\
	\nonumber
	&+\frac{1}{8x}\left[1-x^2\left(\textstyle 1-\frac{i\Omega_m}{\varepsilon_q}\right)^2\right]
	\ln\left(\frac{1+1/x-i\Omega_m/\varepsilon_q}{1-1/x-i\Omega_m/\varepsilon_q}\right)\\
	&\left.-\frac{1}{8x}\left[1-x^2\left(\textstyle 1+\frac{i\Omega_m}{\varepsilon_q}\right)^2\right]
	\ln\left(\frac{1-1/x+i\Omega_m/\varepsilon_q}{1+1/x+i\Omega_m/\varepsilon_q}\right)\right\},
\end{align}
with the static limit
\begin{equation}
	-\frac{e^2}{\epsilon_0}\Pi_0(q,0)=\kTF^2\left(\frac{1}{2}+\frac{1-x^2}{4x}
	\ln\left|\frac{1+x}{1-x}\right|\right).
\end{equation}
We obtain for the stationary interaction
\begin{equation}
	\upsilon^{(\ell)}(0) = \frac{1}{N(\eF)}\int_0^1dx\,x
	\frac{P_{\ell}\left(1-2x^2\right)}
	{\pi\kF\aB x^2+\frac{1}{2}+\frac{1-x^2}{4x}\ln\big|\frac{1+x}{1-x}\big|
	+\frac{\pi a_0 \omega_s^2 (1+4\kF^2 s^2x^2) }{ 4k_F c_s^2}}
	\label{eq:usLHzeq0}
\end{equation}
with $N(\eF)$ the density of states per spin. The interaction for $\ell=0,1,2$ is displayed in Fig.~\ref{fig:angular-momenta} as a function of $r_s$ in the absence of the phonon term (i.e., with $\wpn=0$) as the solid curves and including the phonon term as the dashed curves. We see that the purely electronic $p$-wave interaction has an attractive region for $r_s\gtrsim13$. This is the Kohn--Luttinger mechanism \cite{kohn1965new} where, as a result of the Friedel oscillations of the screened Coulomb interaction, a channel of attraction can open. Regarding the phonon-screened interaction, if we take the limit of the standard jellium model with infinitely compressible charge compensating background ({\it i.e.} taking $c_s\rightarrow 0$) the static interaction including phonons is zero in all angular momentum channels. The effect of a non-zero and positive compressibility of the charge compensating background is to blue-shift the phonons. In this case, the phonons cause only a partial screening of the static Coulomb interaction, as revealed by the phonon screened interaction shown in Fig.~\ref{fig:angular-momenta} for the $s$-wave, $p$-wave and $d$-wave pairing channels.

To unify the Kohn--Luttinger mechanism and electron-phonon coupling within a single theoretical framework, we need to take into consideration the interaction for $\Omega_m\neq 0$. Here one runs into an obstacle. For simplicity, we analyze the case $\wpn=0$, but the problem exists also for $\wpn\neq 0$. Neglecting the phonons in Eq.~\eqref{eq:interaction_ell}, the interaction is
\begin{equation}\label{eq:vl}
	\upsilon_{kp}^{(\ell)}(i\Omega_m) = \frac{1}{N(\eF)}\frac{\kF^2}{kp}
	\int_{\frac{|k-p|}{2\kF}}^{\frac{|k+p|}{2\kF}}
	\frac{dx\,xP_{\ell}\left(1-2x^2\right)}
	{\pi\kF\aB x^2+\left\{\frac{1}{2}+\cdots\right\}},
\end{equation}
where $\left\{\frac{1}{2}+\cdots\right\}$ is the dimensionless part of the polarization written in Eq.~\eqref{eq:Pi0}. If $\Omega_m=0$, this quantity behaves as $1-x^2/3$ for $x\to0$, such that the integral in Eq.~\eqref{eq:vl} is well-behaved at $x=0$ and converges at the lower bound, even on the momentum shell $k=p=\kF$. If $\Omega_m\neq0$, however, $\left\{\frac{1}{2}+\cdots\right\}$ behaves as $x^2/\Omega_m^2$ and the integral is logarithmically divergent at the lower bound if $k=p$. Clearly, we can no longer use the conventional $k=p=\kF$ scheme. The problem is avoided when using the on-energy-shell approximation, as replacing $i\Omega_m$ by $\eF(p^2-k^2)/\kF^2$ suppresses the log-singularity at $k=p$. However, the singularity \emph{does} play a role in the Eliashberg equations. On the other hand, given that the divergence is of the logarithmic variety, the singularity should be integrable in Eq.~\eqref{Eq:SelfEnergy} and most likely the momentum-dependent Eliashberg equations remain well-behaved. 

Treating this problem requires application of the theoretical methods presented in the present paper. For this reason, we believe that the theoretical and numerical methods described here contribute to the unification of electronic and phonon-mediated pairing.

%

\end{document}